\documentclass[11pt]{article}
\usepackage[a4paper,total={7in, 10in}]{geometry}
\usepackage{authblk}
\usepackage{dsfont}
\pdfoutput=1
\usepackage{mathtools}
\usepackage{amsmath}
\usepackage{amssymb}
\usepackage[section]{placeins} \usepackage{lineno}
\title{Pump-probe Dark-field X-ray Microscopy}

\author[1,2,*]{Ishwor Poudyal}
\author[2]{Zhi Qiao}
\author[3]{Arndt Last}
\author[4]{Michael R. Armstrong}
\author[2,*]{Zahir Islam}

\affil[1]{Materials Science Division, Argonne National Laboratory Lemont, IL 60439 (USA)}
\affil[2]{X-ray Science Division, Argonne National Laboratory, Lemont, IL 60439 (USA)}
\affil[3]{Karlsruhe Institute of Technology, Institute of Microstructure Technology, Hermann-von-Helmholtz-Platz~1, Eggenstein-Leopoldshafen D-76344 (Germany)}
\affil[4]{Physical and Life Sciences, Lawrence Livermore National Laboratory, Livermore, CA 94550, (USA)}
\affil[*]{Corresponding author: ipoudyal@anl.gov, zahir@anl.gov }

\date{}

\begin{document}

\maketitle
\begin{abstract}

A pump-probe dark-field X-ray microscopy (DFXM) experiment was carried out at the Advanced Photon Source (APS) with nanosecond drive laser pulses and hybrid mode X-ray probe pulses. We observe a thermal decay due to laser-induced heat diffusion in a Germanium single crystal which matches the theoretical prediction. The single pulse DFXM imaging in combination with the laser-pump X-ray probe method could reveal thermal strain formation and propagation of the acoustic wave “on-the-fly” generated by a laser-induced lattice deformation. This will open a new avenue of materials research on laser-induced dynamic phenomena in solids at sub-nanosecond time scales.

\end{abstract}

\section{Introduction}
\label{S:intro}
\smallskip
Materials undergo several non-equilibrium minima across the energy landscape as a function of time against external perturbations like temperature, pressure, and stress~\cite{wales2003energy}. These non-equilibrium minima do not exist under steady-state conditions. Understanding these non-equilibrium processes helps to determine the structure, dynamics, and thermodynamics of a given system and also provides the conceptual and computational framework for condensed matter science~\cite{wales2003energy}. However, it remains challenging to systematically drive materials into these states for detailed study.  The studies from the ground-state targets to excited states are enabled by the laser-induced pump-probe experiments.  Other perturbations such as electric field~\cite{karpov2017three} and mechanical loading~\cite{dupraz20173d} have also been used to induce disorder in the materials. Observing and quantifying the response of materials like phase transitions and lattice distortion to external stimuli would shed new light on their behavior. 

Pump-probe measurements have been used to explore the dynamic properties of the interaction of materials, such as interfaces, defects, and surfaces on material properties~\cite{clark2013ultrafast}. This technique has been used extensively for data collection in both solid-state materials~\cite{clark2013ultrafast} and biological materials~\cite{malla2022transient,pandey2020pump}. A pump pulse (laser pulse) initiates dynamics and after a controllable time delay, a probe pulse (X-rays) is used to characterize the resultant change. The temporal resolution is defined by the duration of the pump and probe pulses. For a nondestructive experiment and weak diffraction signal from a single probe-pulse, pump-probe cycles can be repeated many times for a given fixed time delay to increase the signal-to-noise ratio. Yet, this repetitive measurement only allows the characterization of reversible processes. For destructive single-shot pump-probe experiments of non-reversible phenomena (e.g. diffractive imaging of proteins at X-ray Free-Electron Lasers), the sample must be replenished for each probe pulse~\cite{pandey2020time,schmidt2021macromolecular}. 

The study of dynamic strain events with the pump-probe experimental techniques may help to understand fundamental dissipation and transformation mechanisms in materials undergoing inelastic deformation. The inelastic transformation processes, at low strain rates, have been studied using transmission electron microscopy (TEM) but are limited to sample thickness to 100 nm scales. Nano-scale material phenomena are difficult to study with TEM owing to short time scale phenomena at the length scale of material defects~\cite{kondo2016direct}. Even the highest time resolution dynamic TEM cannot resolve acoustic wave modulated nm-scale material transformations due to the blurring from the strain wave motion~\cite{armstrong2007practical}. Understanding fundamental phenomena in materials undergoing transformations (such as dislocation migration in plasticity) will require nm-scale spatial resolution. Dark-field X-ray microscopy (DFXM)~\cite{simons2015dark,simons2018long,yildirim2020probing}, a material characterization tool that emerged in the past decade, can potentially help to obtain the necessary spatial resolution \cite{poulsen2021geometrical} for understanding heterogeneous phenomena in materials. DFXM is a full-field X-ray imaging technique and has been established as a tool for mapping orientation, and strain in deeply embedded structures~\cite{simons2015dark}. This technique has also been employed for studying in situ dynamic processes such as dislocation movement as a function of temperature~\cite{dresselhaus2021situ} and structural transformations during phase transitions in ferroelectric materias~\cite{ormstrup2020imaging}. However, the pump-probe laser scheme incorporating the DFXM technique at synchrotron sources has not yet been reported. Here, we demonstrate the feasibility of combining DFXM with pump-probe experiments at Sector 6 at APS (Advanced Photon Source).
\smallskip 

 The experiments reported here examine the laser-induced strains near the surface of a Germanium single crystal in low X-ray energy and non-destructive mode. Germanium is a popular material for high-speed metal-oxide-semiconductor transistors and silicon-based optoelectronics~\cite{armand2019lasing}.  It has good optical properties and provides carrier mobility much higher than that of silicon \cite{armand2019lasing,rathore2021evolution}. The study of time-resolved full-field X-ray imaging, which can resolve features within the beam size, of strain propagation in Germanium semiconductor could be important for the high-speed device application purposes. A similar experimental approach could be used for direct imaging of propagation of short pulse strain waves, due to laser heating, through the bulk single crystal. These types of studies will inform models of laser-induced electron-phonon interactions which will ultimately provide comprehensive information and underlying mechanisms about laser-induced melting processes in semiconductors.

\section{Experimental Setup}
\label{S:setup}
\begin{figure}[htbp]
	\centering
	\includegraphics[width=0.95\textwidth]{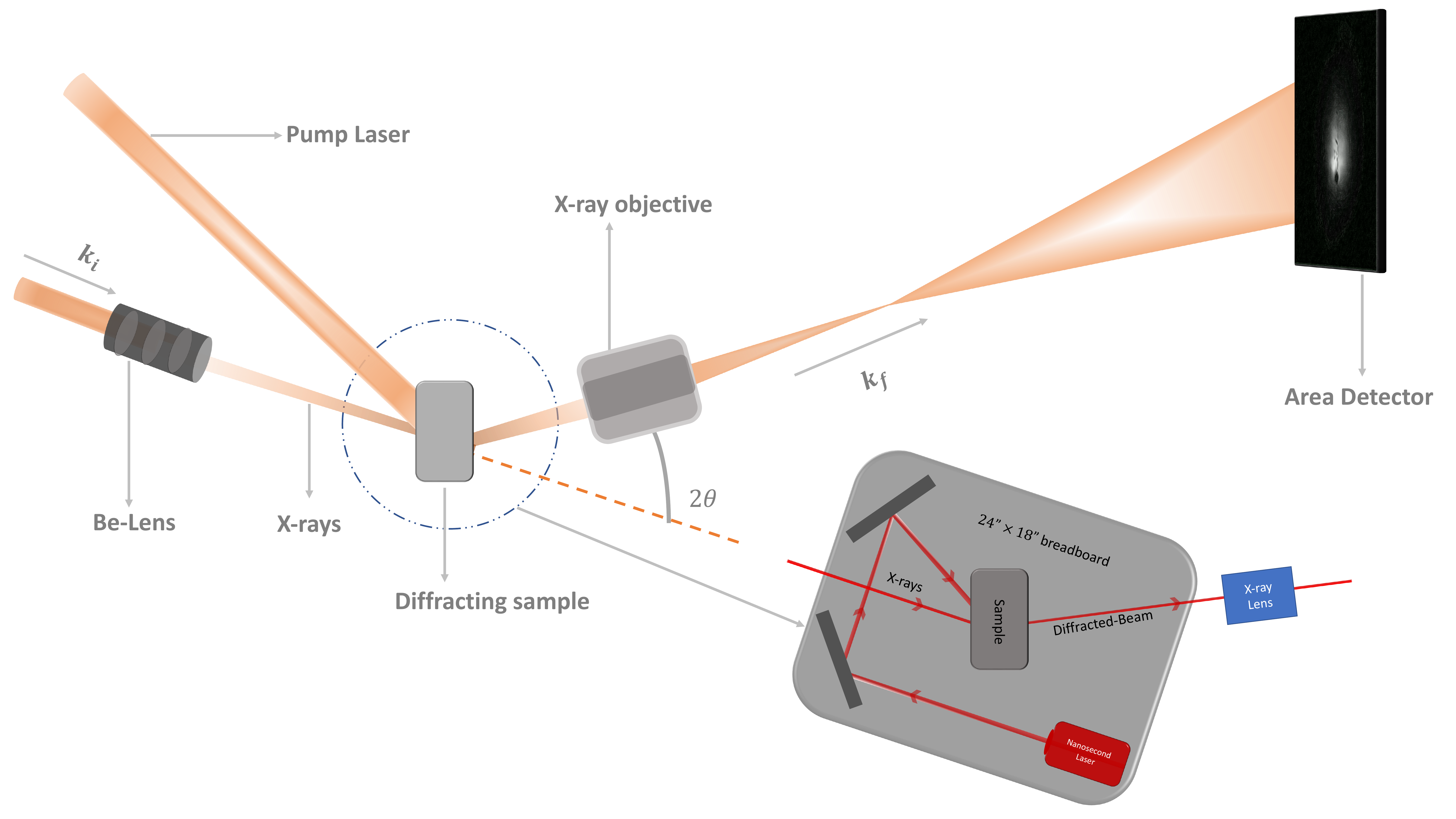}
	\caption{{\textbf{(a)} Schematic of a pump-probe DFXM experiment. The nanosecond laser was mounted on the breadboard as higlighted on the figure which was then bolted on the diffractromter for the measurement.}}
	\label{fig.schematic}
\end{figure}

\begin{figure}[ht!]
	\centering
	\includegraphics[width=0.80\textwidth]{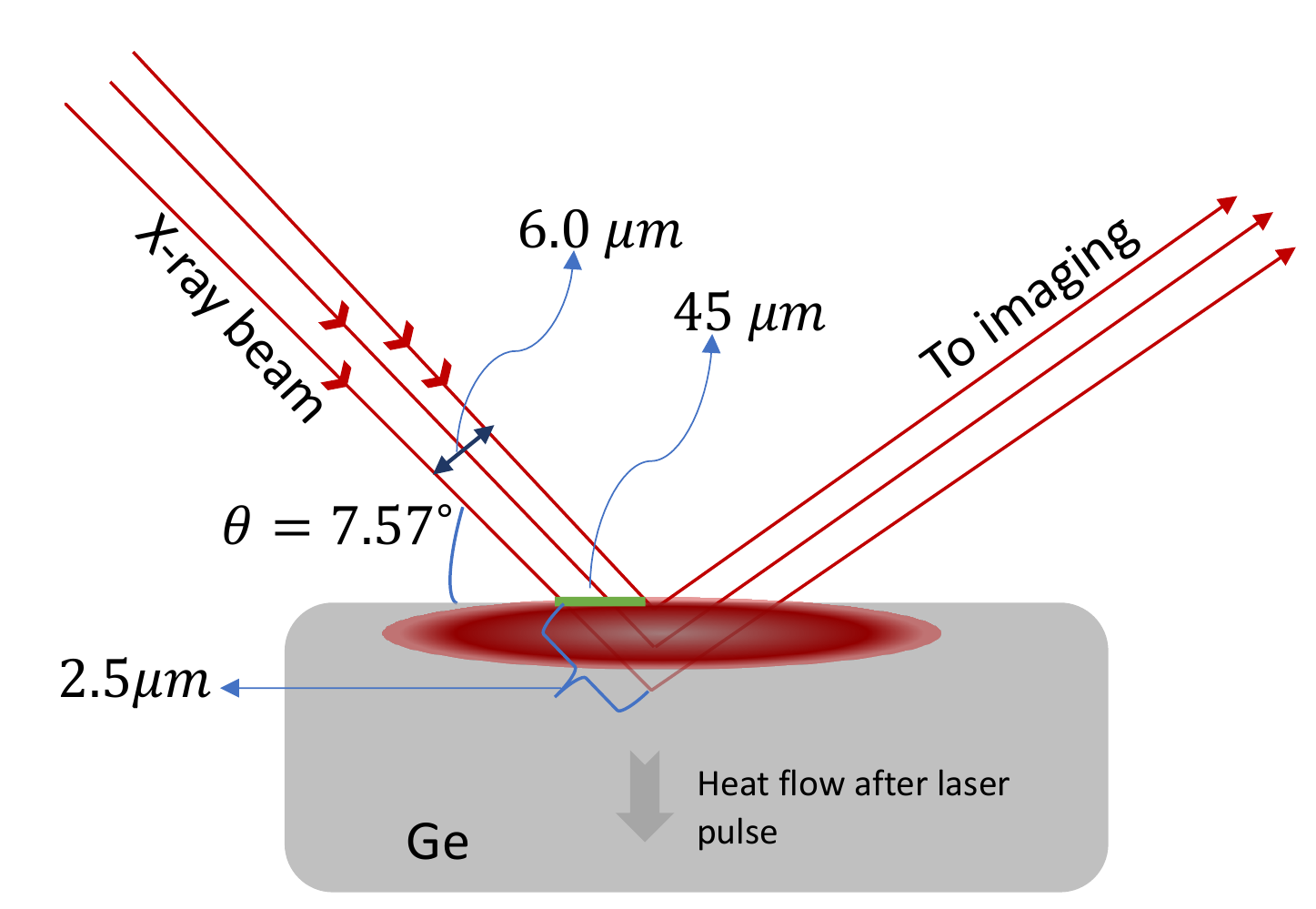}
	\caption{A schematic diagram illustrating the various length scales of the measurement. The reddish blob represents the area illuminated by the laser pulse. The direction of heat flow after the laser pulse is shown by an arrow pointing downwards. The incident-focused X-ray beam has a dimension of $6.0~\mu m$ near the sample which produced a footprint of $45~\mu m$  on the sample in the horizontal direction (solid green line). The absorption depth of the Germanium is $2.5~\mu m$ at the energy of 13~keV so we only see the effect on the surface of the Germanium during measurements. }
	\label{fig.schematic_lengthscale}
\end{figure}

\smallskip
A schematic of the DFXM geometry for this experiment is shown in Fig.(\ref{fig.schematic}). Our experiments were carried out at Beamline 6-IDC at the Advanced Photon Source~\cite{qiao2020large}. An X-ray beam with an energy of $13.0~$keV selected by a Si (111) double crystal monochromator, with a bandwidth of $\Delta E/E=10^{-3}$. The beam was condensed using a Be Compound Refractive Lens (CRL) comprised of eight 2D Be lenslets, with a radius of curvature $R=50~\mu m$ that produces an effective focal length of $1.5~m$, to generate a focused beam of size $6~\mu m \times 6~\mu m$ on the sample. A $5~mm \times 5~mm \times 0.5~mm$ single crystal Germanium sample is mounted on the high-precision translation and rotation stages such that the (111) Bragg peak is measured in the reflection geometry in the horizontal scattering plane. Because of the reflection geometry, there is a lower axial resolution (more blurring along the beam-direction due to the footprint on the sample (see Fig.(\ref{fig.schematic_lengthscale}))) in our measured images. The transverse spatial resolution is given by the focused-beam size produced by the condenser lens and is roughly equal to one half of the focused-beam size. The single-crystalline nature of the sample was verified through static X-ray diffraction measurements carried out at the same beamline during the same experiment, which shows the capability of switching the setup for both diffraction and imaging experiments.  The 10~ns (FWHM) pulsed laser (1064~nm wavelength, Quantel-ICE450) with maximum repetition rate of 10~Hz was used to induce surface heating of the Ge crystal, as shown in Fig.(\ref{fig.schematic}).The X-ray attenuation length of Germanium is small at 13 keV energy and only the effect on the surface of the Germanium is visible during measurements as shown in Fig(\ref{fig.schematic_lengthscale}). The excited sample is probed after a fixed time delay with the X-ray pulse. The direction of the diffracted beam in the horizontal plane is characterized by the scattering angle for a nominal reflection (111). The objective lens, polymeric CRL (pCRL)\footnote{Layout 1921\_00\_A0 \#18, lot 2018-04\_400-6.}, is aligned so that the optical axis lies along the diffracted beam to produce a magnified image on the 2D Pi-MAX IV detector. The objective lens has a design focal length of $\sim$ 95~mm at 13~keV with an effective aperture of $\sim 60~\mu m$.  The detector housing is coupled with a scintillator, which converts X-ray photons to visible light, and the optical setup with $M_o=5\times$ magnification \cite{qiao2020large}.  The objective lens projected a magnified image of the diffracting sample onto the far-field detector, with an X-ray magnification of $M_{X-ray}=19.0\times$. This yields a total magnification of $M_t=95\times$.
\smallskip 

\begin{figure}[htbp]
	\centering
	\includegraphics[width=0.80\textwidth]{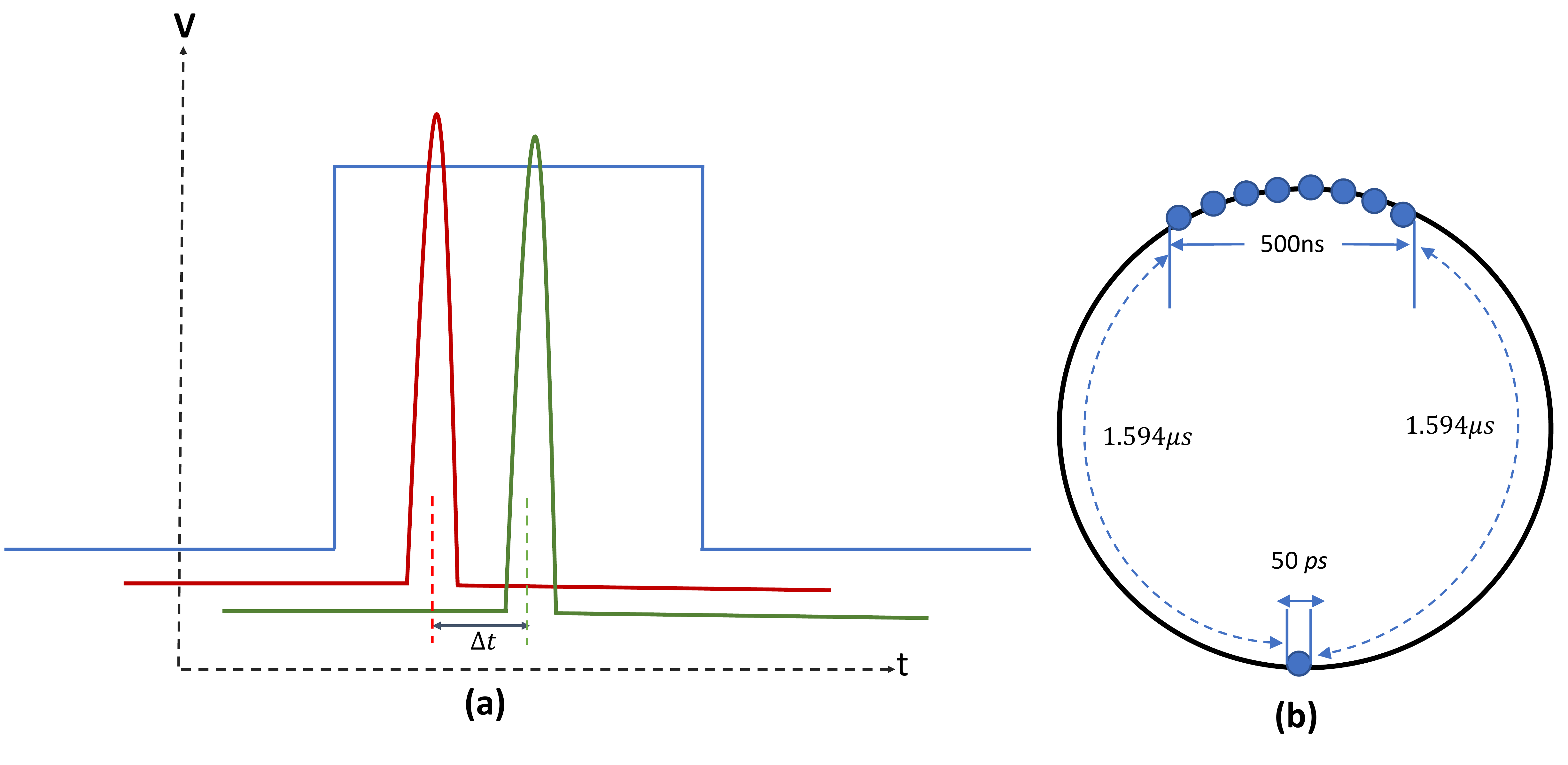}
	\caption{\textbf{(a)} Schematic of the pump-probe time delay as visualized on the oscilloscope. The blue rectangular pulse represents the gatewidth of the PI-MAX detector. The green and red pulses represent the laser (pump) pulse and the single-bunch X-ray (probe) pulse respectively. The time delay ($\Delta t$) is the time-delay between the pump and the probe pulse.
	\textbf{(b)} Schematic of the special hybrid operating mode (hybrid fill, top-up) of the APS Storage Ring. A single bunch is isolated from the remaining bunches by symmetrical $1.594~\mu s$ gaps. The singlet has a bunch length of 50~ps with the total length of the remaining bunch train being $500~ns$. }
	\label{fig.schematic_timing}
\end{figure}

\smallskip
The experiment was conducted when the APS storage ring was operating in hybrid mode. In hybrid mode, a single bunch is diametrically opposed to eight septuplets, as shown in Fig.(\ref{fig.schematic_timing}). The gap between the isolated bunch and the closest bunch in the septuplets is $1.59~\mu s$, with the length of the bunch of each singlet equal to 50~ps. The normal operating mode has 24 bunches circulating with 153~ns separation between the bunches. The 10~ns duration excitation laser was used to excite the sample. The pump laser was 5~mm in diameter and the laser illumination was uniform over the X-ray beam profile. To synchronize the X-ray and the laser pulse, we placed a photodiode near the sample position and recorded the laser pulses. The avalanche photodiode detector (APD) was installed at 45 degrees from the path of X-rays to detect the signal from the X-ray pulse. The signals from the APD and the photodiode were observed on a 1 GHz oscilloscope as shown in Fig.(\ref{fig.schematic_timing}). The timing between the pump and the probe pulse was adjusted with the DG635 delay generator and monitored with the oscilloscope. In this way X-rays were used to probe the laser-excited sample volume at a precisely controlled and variable time delay.
\smallskip

For these types of time-resolved measurements, the imaging detector should differentiate between individual X-ray bunches emitted by the synchrotron. A Princeton Instruments PI-MAX4 detector was used for this purpose as an imaging detector for our experiment. The detector gate delay, which is an internal time delay generator of the camera, with respect to the nanosecond pump pulse, as well as the integration period (gate width), was chosen so that the camera is exposed to scintillation light from only a single X-ray pulse. The PI-MAX4 camera has a $13~\mu m \times 13~\mu m$ pixel size and a detection area of 13~mm $\times$ 13~mm or 1024 $\times$ 1024 pixels. The camera was triggered externally by DG635 and can record only one frame when triggered.  Fig.(\ref{fig.integratedImages}) shows an example of static X-ray images obtained with a PI-MAX4 camera (a signal gain of 100$\times$). Both static and dynamic images were collected using the same PI-MAX detector.

\section{Results}
\label{S:results}
\begin{figure}[htbp]
	\centering
	\includegraphics[width=1.0\textwidth]{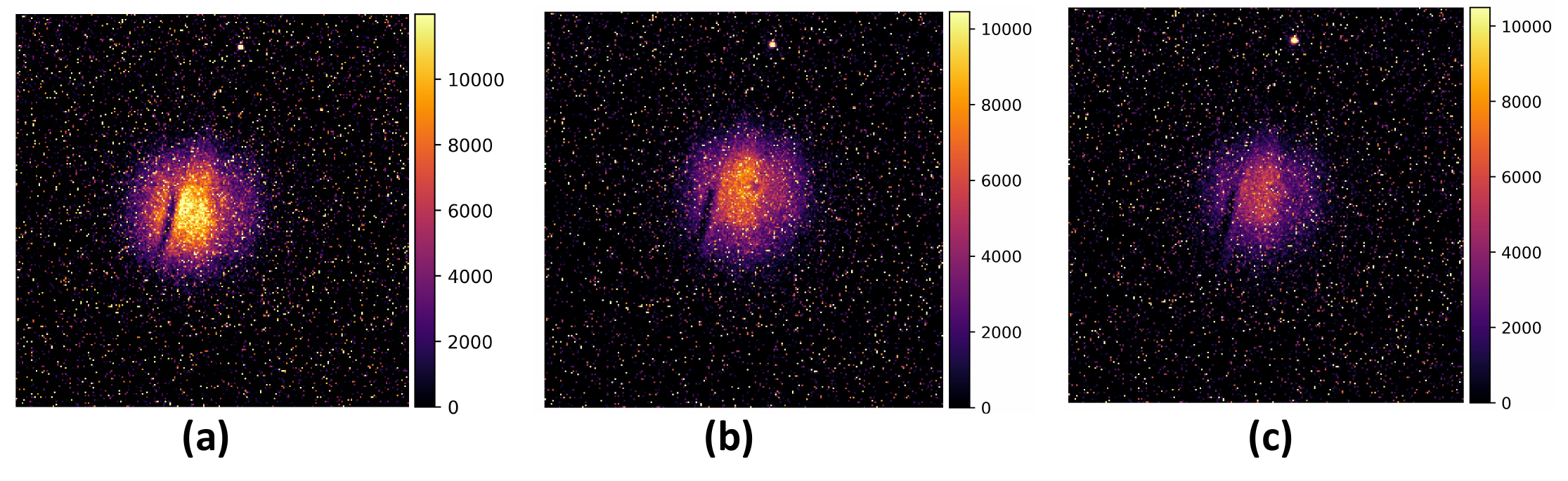}
	\caption{{DFXM images showing the integrated intensity (over 1000 single shots). \textbf{(a)} An integrated static image of the Germanium crystal, i.e., without any laser pulse. \textbf{(b) (c)} Integrated images after 15~ns and 60~ns after the application of laser pulse respectively. The color code corresponds to the integrated intensity in arbitrary units. The decay of integrated intensity is visible in these images.}}
	\label{fig.integratedImages}
\end{figure}

\begin{figure}[htbp]
	\centering
	\includegraphics[width=0.95\textwidth]{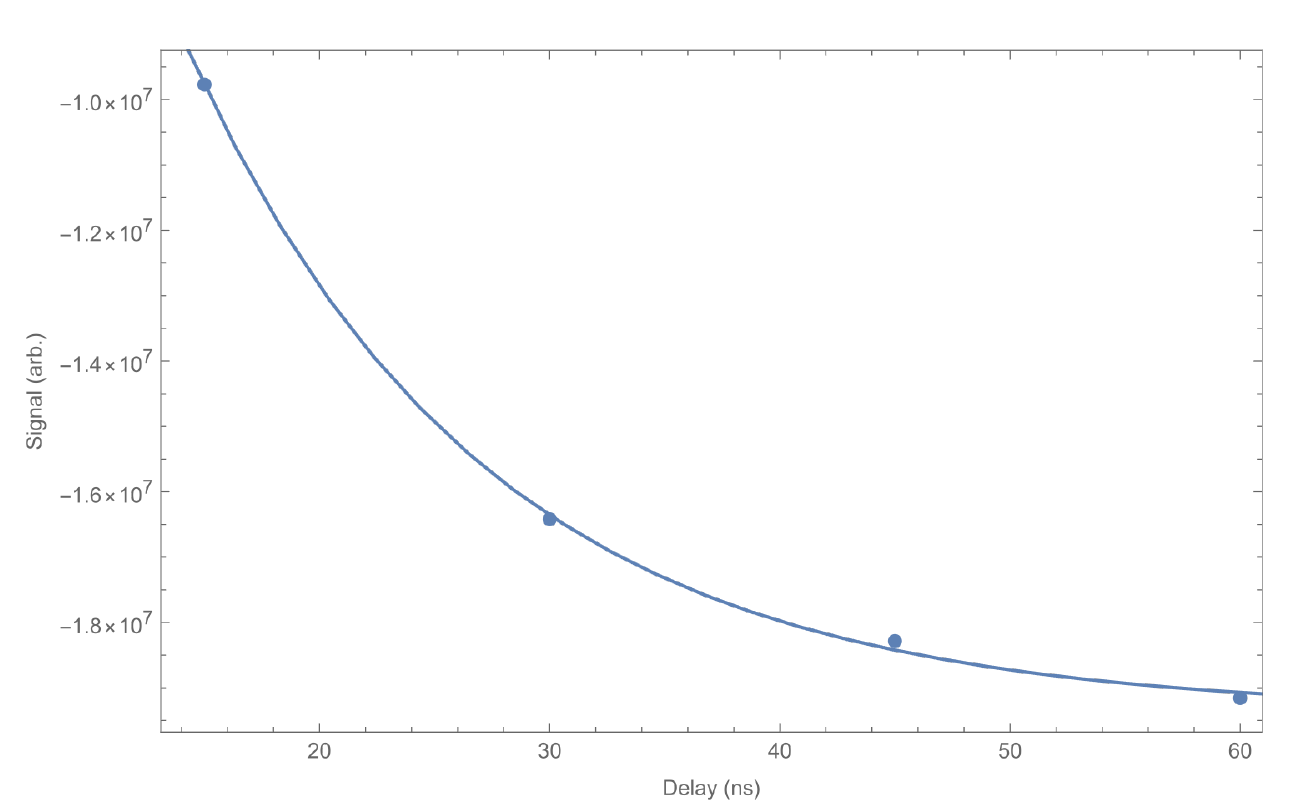}
	\caption{{Plot of integrated intensity as a function of delay time. The solid circles represent the experimental data whereas the solid line corresponds to a fit to the data assuming an exponential decay.}}
	\label{fig.exponentialfit}
\end{figure}

\smallskip
The static image and the images recorded for different time delays are shown in Fig.(\ref{fig.integratedImages}). The experiment was repeated without moving the sample and the signal was integrated over 1000 shots for a better signal-to-noise ratio for dynamic images. The images were recorded at various time delays (i.e., probe delays) after the excitation of the sample with the drive laser pulse. The width of the drive pulse is much larger than the probe pulse, so the laser illumination can be assumed to be uniform over the X-ray beam profile. The intensity of X-ray diffraction decreases as a function of delay time, as shown in Fig.(\ref{fig.exponentialfit}), resulting from the heterogeneous strain introduced into the sample. The moderate decrease in diffraction intensity can be attributed to the combination of shifting of the scattering angle with respect to the aperture of the pCRL and  the Debye-Waller factor, $e^{\frac{-1}{2}Q^2<u^2>} $, where $<u^2>$ denotes the mean-square displacement of collective atoms along the scattering vector $Q$~\cite{als2011elements}. Also, the decay of intensity is evidence that the Germanium lattice is slightly strained as a result of laser-induced heating. The laser fluence for our experiment was not sufficient enough to cause surface melting.  If the laser fluence is lower than the damage threshold, the crystal expands due to lattice heating via carrier diffusion~\cite{decamp2005x}, or strain is produced on the crystal surface due to compression and expansion during the propagation of photo-induced acoustic waves~\cite{thomsen1986surface}. For a nanosecond time scale measurement, the laser pulse duration is longer than the time scale of electron-phonon coupling, which happens over picoseconds. On these time scales, the electrons and lattice temperatures will remain at thermal equilibrium~\cite{hamad2016effects}.  The repetitive laser exposure, in our experiment, strained the crystal and the average spacing of the crystal lattice changed accordingly. For our measurements, we rock the crystal to measure the intensity distribution as a function of the rocking angle. The integrated intensity as shown in Fig.(\ref{fig.integratedImages}) is the contribution of the X-ray intensity only from the perturbed crystals and is obtained from the difference between the perturbed and unperturbed intensity distributions.
\smallskip

The exponential decay is a standard representation of the statistical decays for thermally activated processes~\cite{langer1969statistical}. The general shape of the integrated X-ray intensity, measured during the experiment, as a function of the delay time is well described by an exponential function of the form $I=A+Be^{-(t/\tau) }$ where $t$ is the delay time and $\tau$ is the thermal decay time. The phenomenological exponential decay model shown as a solid line in Fig.(\ref{fig.exponentialfit}), provides a good description of our measured data. Based on this model, the thermal decay time was estimated to be 12.94~ns with an error of $1\%$. The error corresponds to the calculated thermal decay time based on known absorption depth for X-ray energy at 13 keV and diffusion constant calculated from thermal conductivity and heat capacity of the Germanium crystal. 

\section{Conclusions}
\label{S:conclusions}

\smallskip
We have established the feasibility of pump-probe DFXM experiments at APS. We studied non-destructive laser-induced strain formation near the surface of a Germanium single crystal induced by a nanosecond optical laser pulse excitation using the DFXM technique.  Our results substantiate the potential of the pump-probe DFXM experiment to address a variety of material properties under conditions far from room temperature and pressure \cite{radousky2020melting,radousky2021time}. The use of lasers with DFXM has given access to the new research field and will probably answer many scientific queries with synchrotron radiation. With the upgraded synchrotron sources like APS-U~\cite{henderson2015status}, the nature of the experiments and the quality of the results will also improve dramatically, especially when lasers are combined with synchrotron radiation beams. Direct visualization of the evolution of laser-induced subsurface heating and melting, propagation of the strain waves in the materials~\cite{holstad2022x} at synchrotron sources will become possible in the future by combining the knowledge gained from this experiment. 

\section{Acknowledgements}

This research used resources of the Advanced Photon Source, a U.S. Department of Energy (DOE) Office of Science user facility at Argonne National Laboratory and is based on research supported by the U.S. DOE Office of Science-Basic Energy Sciences, under Contract No. DE-AC02-06CH11357. I.P is supported by Laboratory Directed Research and Development (LDRD) funding from Argonne National Laboratory, provided by the Director, Office of Science, of the U.S. DOE under Contract No. DE-AC02-06CH11357. The authors would also like to thank the Karlsruhe Nano Micro Facility (KNMF) for the fabrication of the polymer X-ray optics.
\smallskip

\clearpage



\bibliographystyle{unsrt}
\bibliography{main.bib}

\end{document}